\documentclass{article}

\usepackage{arxiv2}

\usepackage[utf8]{inputenc} 
\usepackage[T1]{fontenc}    
\usepackage{hyperref}       
\usepackage{url}            
\usepackage{booktabs}       
\usepackage{amsfonts}       
\usepackage{nicefrac}       
\usepackage{microtype}      
\usepackage[utf8]{inputenc}
\usepackage{lipsum}		
\usepackage{graphicx}
\usepackage{natbib}
\usepackage{doi}
\usepackage{authblk}
\usepackage{subcaption}
\usepackage{multirow}
\usepackage{bm}
\usepackage{siunitx}
\usepackage{soul}
\usepackage{amsmath}

\title{Improving medium-range ensemble weather forecasts with hierarchical ensemble transformers}

\author[1]{Zied {Ben Bouall\`egue}}
\author[1,2]{Jonathan A Weyn}
\author[1]{Mariana C A Clare}
\author[1]{Jesper Dramsch}
\author[1]{Peter Dueben}
\author[1]{Matthew~Chantry}
\affil[1]{ECMWF}
\affil[2]{Microsoft}

\renewcommand{\shorttitle}{Post-processing of Ensembles with Transformers}

\hypersetup{
pdftitle={A template for the arxiv style},
pdfsubject={q-bio.NC, q-bio.QM},
pdfauthor={David S.~Hippocampus, Elias D.~Striatum},
pdfkeywords={First keyword, Second keyword, More},
}

\begin{document}
\maketitle

\begin{abstract}
Statistical post-processing of global ensemble weather forecasts is revisited by leveraging recent developments in machine learning. Verification of past forecasts is exploited to learn systematic deficiencies of numerical weather predictions in order to boost post-processed forecast performance. Here, we introduce PoET, a post-processing approach based on hierarchical transformers. PoET has 2 major characteristics: 1) the post-processing is applied directly to the ensemble members rather than to a predictive distribution or a functional of it, and  2) the method is ensemble-size agnostic in the sense that the number of ensemble members in training and inference mode can differ. The PoET output is a set of calibrated members that has the same size as the original ensemble but with improved reliability. Performance assessments show that PoET can bring up to 20\% improvement in skill globally for 2m temperature and 2\% for precipitation forecasts and outperforms the simpler statistical member-by-member method, used here as a competitive benchmark. PoET is also applied to the ENS10 benchmark dataset for ensemble post-processing and provides better results when compared to other deep learning solutions that are evaluated for most parameters. Furthermore, because each ensemble member is calibrated separately, downstream applications should directly benefit from the improvement made on the ensemble forecast with post-processing. 
\end{abstract}


\section{Introduction}

The chaotic nature of the atmosphere makes forecasting the weather a challenging and scientifically exciting task.
With large and high-quality publicly available datasets \citep{era5}, weather forecasting is becoming a new playing field for deep-learning practitioners \citep{fourcastnet,pangu,graphcast}. More traditionally, at national meteorological centres, weather forecasts are generated by numerical weather prediction (NWP) models that resolve numerically physics-based equations. A Monte-Carlo approach is followed to account for uncertainties: an ensemble of deterministic forecasts is run with variations in the initial conditions, the model parametrizations, and/or the numerical discretization. This ensemble approach, initially developed to explore the limits of deterministic forecasting, has now become the backbone of operational weather forecasting \citep{lewis05}. 

Practically, ensemble weather forecasts are a set of physically consistent weather scenarios that ideally capture the full range of possible outcomes given the information available at the start of the forecast \citep{leutbecher2008ensemble} and decision-making can be optimized using probabilistic forecasts derived from such an ensemble \citep{richardson2000}. One can assess not only the uncertainty of a weather variable at a given point in space and time, but also any joint probability distributions across the variables. This versatility is essential for ensemble prediction systems to support downstream applications with high societal relevance, such as flood forecasting or human heat stress forecasting \citep{magnusson2023}.

However, as an output of a NWP model, an ensemble forecast is sub-optimal in a statistical sense. On top of the limited ensemble size effect \citep{leutbecher2019ensemble}, systematic deficiencies like model biases (defined as the averaged differences between forecasts and observations) and over- or under-dispersiveness (too much or too little ensemble spread  as measured by the standard deviation among the ensemble members) are common features of any NWP ensemble forecasts \citep{haiden2021}. Statistical post-processing is proposed as a simple remedy where past data is exploited to learn forecast errors and correct the current forecast accordingly. 

A variety of post-processing approaches have been used over the years, from simple bias correction to machine learning based methods \citep[][]{vannitsem2021statistical}. Classically, post-processing of ensemble forecasts is achieved either by assuming the form of the predictive probability distribution and optimizing its parameters \citep{gneitingCalibratedProbabilisticForecasting2005,raftery2005} or by correcting a limited set of quantiles of the predictive distribution \citep{taillardatCalibratedEnsembleForecasts2016,zbbmetzet2016}. Recently, multiple different modern machine learning methods have been applied to ensemble post-processing. For example, \citet{raspNeuralNetworksPostprocessing2018} trained a neural network to predict mean and standard deviation of a normal distribution for 2m temperature forecasting at stations in Germany while \cite{bremnesEnsemblePostprocessingUsing2020} combined a neural network and Bernstein polynomials for generating quantile forecasts of wind speed at stations in Norway. In the case of downstream applications based on such a post-processed forecast, an additional post-processing step is required to ``reconstruct'' forecast dependencies between variables or in time and space \citep{decc2015,baran2020}. However, more recent developments in deep learning, particularly transformers, promise to resolve this issue by using mechanisms such as attention \citep{Vaswani2017} to maintain inter-variable and inter-spatial dependencies.

In this work, we target the direct generation of a calibrated ensemble for potential use in downstream applications. The focus is on 2m temperature and precipitation, which are variables of interest for many stakeholders. We propose a new approach for ensemble post-processing: \textit{PoET} (Post-processing of Ensembles with Transformers). Our machine learning framework, PoET, combines the self-attention ensemble transformer used for post-processing of individual ensemble members in \citet{Finn2021} with the U-Net architecture used for bias correction in \citet{Grnquist2021}, leveraging the advantages of both for the first time in a post-processing application. We compare this approach with the statistical member-by-member (MBM) method proposed by \citet{VanSchaeybroeck2015} that is simpler thanPoET but effective. In its simplest form, this method consists of a bias correction and a spread scaling with respect to the ensemble mean. MBM has been successfully tested on time series of 2m temperature ensemble forecasts and is now run operationally at the Royal Meteorological Institute of Belgium \citep{demaeyer2021}. 

Machine learning approaches for ensemble post-processing rely on the availability of suitable datasets \citep{dueben2022challenges}. Here, we use reforecasts and reanalysis for training. In this context, reforecasts and reanalysis are praised for their consistency because they are generated from a single NWP model for long periods of validity time. In particular, the benefit of reforecasts for post-processing has been demonstrated in pioneering works on 2m temperature and precipitation forecasts at station locations by \cite{hagedorn2007} and \cite{hamill2007}, respectively. Reforecasts are also becoming the cornerstones of benchmark datasets for post-processing of weather forecasts \citep{ens10,euppbench2023,Grnquist2021}. In this work, we continue this trend focusing on ensemble post-processing of global gridded forecasts.

The remainder of this paper is organized as follows: Section \ref{sec:methods} introduces the dataset and methods investigated in this study;  Section \ref{sec:experiments} provides details about the implementation of MBM and PoET for the post-processing of 2m temperature and precipitation ensemble forecasts as well as a description of the verification process; Section \ref{sec:examples} provides illustrative examples of post-processing in action; Section \ref{sec:results} presents and discusses verification results and Section \ref{sec:conclusion} concludes this paper.

\section{Data and Methods}
\label{sec:methods}
\subsection{Data}
\label{sec:data}

At the European Centre for Medium-Range Weather Forecasting (ECMWF), the reforecast dataset consists of 11 ensemble members (10 perturbed + 1 control) generated twice a week over the past 20 years \citep{vitart2019}. In our experiments, the dataset comes from the operational Integrated Forecasting System (IFS) reforecasts produced in 2020, that is with IFS cycles 46r1 and 47r1, with the switch in June 2020. Fields are on  1 degree horizontal grid resolution and the focus is on lead times every 6h up to 96h. Reforecasts from 2000 to 2016 are used for training, while those in 2017 and 2018 are used for validation.

The post-processing models are trained towards ERA5, the ECMWF reanalysis dataset \citep{era5}. The target is the reanalysis of 2m temperature while the short range forecasts at T+6h (aligned with the forecast validity time) is used as a target for precipitation to account for the spin-up after data assimilation \citep[for a comprehensive assessment of ERA5 daily precipitation please refer to][]{lavers2022}. 

For testing, we use the operational ensemble data from 2021, using two forecasts each week for 104 start dates in total, according to the ECMWF sub-seasonal-to-seasonal (S2S) model iterations. The operational ensemble has 51 members (50 perturbed members + 1 control member), but we apply post-processing methods that are agnostic to ensemble size: they may be run in inference mode with a different ensemble size than used in training.  The data from 2021 includes model cycles Cy47r1, Cy47r2 and Cy47r3, switching in May and then October of 2021, respectively. Notably the model upgrade in Cy47r2 included an increase to 137 vertical levels in the ensemble, an improvement that is not included in the training dataset. We are therefore directly testing our methodology for generalization across model cycles, an important property to reduce the maintenance required when operationalizing machine learning systems.

\subsection{Statistical benchmark method for comparison}
\label{sec:benchmark}
Neural networks are not the only methods that can be used to calibrate ensembles. There exist simpler statistical methods, which require less computational power and which are generally more \textquoteleft explainable\textquoteright. In this work, we use the member-by-member (MBM) approach detailed in \cite{VanSchaeybroeck2015} as a benchmark. MBM is a natural benchmark for PoET that can be seen as a sophisticated member-by-member method. In addition, a comparison with state-of-the-art ML post-processing techniques is discussed in the framework of the \textit{ENS-10} benchmark dataset \citep{ens10} in Section \ref{sec:ens10}. 

With MBM, a correction is applied to each ensemble member individually with a component common to all members and a component that adjusts the deviation of a member with respect to the ensemble mean.  Let's denote $\hat{x}_{i}$ the corrected forecast for the $m^{\text{th}}$ member of the ensemble. Formally, MBM consists of applying:
\begin{equation}
 \hat{x}_{i} = \alpha + \beta\overline{x} + \gamma(x_i - \overline{x}),
\end{equation}
where $x_i$ is the ensemble member $i$ and $\overline{x}$ the ensemble mean. The parameter $\alpha$ is the bias parameter that nudges the ensemble mean, $\beta$ is the linear coefficient that scales the ensemble mean, and $\gamma$ is the scaling parameter that adjusts the spread of the ensemble. Each parameter can be inspected separately to understand their respective contribution to the modifications of the forecasts. 

In our application, the parameters optimization follows the so-called \textit{WER+CR} approach as defined in \cite{VanSchaeybroeck2015}. \textit{WER+CR} means that the estimated parameters are constrained to preserve two different reliability conditions, the \textit{WER} and the \textit{CR} conditions. For bias-free forecasts, climatological reliability (\textit{CR}) is defined as the equality of forecast variability with observations variability, while weak ensemble reliability (\textit{WER}) is defined as the agreement between average ensemble variance and the mean squared forecast error. The analytical formulae used to compute the 3 MBM parameters are provided in Appendix \ref{sec:mbmopt}). Note that other flavors of MBM exist (\textit{e.g.} with score optimization), but they have been disregarded because of their prohibitive computational costs in our application. For example, the MBM approach based on the minimization of the continuous ranked probability score (the so-called \textit{CRPS MIN} approach) is 3 orders of magnitude more computationally expensive. Furthermore, a test of the \textit{CRPS MIN} approach on a sub-sample of the data shows no benefits in terms of scores compared with the method applied here.

\subsection{The ensemble transformer}
\label{sec:transformers}

Transformers are a class of neural networks that were designed for large natural language processing (NLP) models \citep{Vaswani2017}. The main advantage of transformers is the capability to process arbitrary lengths of sequences, drawing context from every part of the sequence, without the expensive sequential computations and potential saturating gradient issues of recurrent methods such as long short-term memory (LSTM; \cite{Hochreiter1997}). Deep-learning transformer architectures are often structured as encoder-decoder networks, where the encoder blocks use a self-attention layer to compute correlations between all elements of an input sequence.

Self-attention uses a key-query-value construction which shares some similarities with Kalman filters, often used in meteorology for such tasks as data assimilation. The learnable parameters are weight matrices $W^K_l, W^Q_l, W^V_l$ that encode the response to input tensors $X_l$ at the $l$-th layer such that $K = W^K_l X_l$, $Q = W^Q_l X_l$, and $V = W^V_l X_l$ are the key, query, and value, respectively. Note that the term \textquoteleft self-attention\textquoteright \text{ }refers to the operation of each of these components on the same input latent state tensor $X_l$. The resulting scaled dot-product attention for layer $l$ is

\begin{equation}
\textbf{Attention}_l(Q_l,K_l,V_l)=\textbf{softmax}\left(\frac{Q_l K_l^T}{\sqrt{d_k}}\right)\cdot V_l    
\label{eq:attention}
\end{equation}

where $d_k$ is the dimensionality of keys and queries. 
In implementation, the weights are $1\times1$ convolution layers (i.e., dense layers) with a small number of filters, or heads, which produce multiple attention maps, a process known as multi-head attention. 


While in NLP models, the attention is computed along the dimension of the encoded input sequence, \citet{Finn2021} proposed applying this methodology across the ensemble dimension for ensemble NWP forecasts.
The strategy has plenty of appeal: by computing the similarity between ensemble members, one can dynamically integrate information from the complete ensemble to post-process each individual member. 
Consider an input tensor of shape $B\times N \times C \times H \times W$, where the dimensions represent the batch size, number of ensemble members, number of feature channels (such as different atmospheric variables), and the height and width of the model grid, in latitudes and longitudes.
This transformer is applied along the ensemble dimension of the forecasts, resulting in a re-weighting tensor of shape $B \times K \times N \times N$, where $K$ is the number of attention heads.
The weights are computed from the scalar dot product (Eq.~\ref{eq:attention}) over the channel, height, and width dimensions.
This particular implementation makes the transformer scalable to different spatial input dimensions from global to regional models and is also agnostic to the number of ensemble members, meaning that the model can be trained on a limited number of ensemble members but inference can be run on a larger ensemble.
Taking inspiration from ensemble data assimilation, the attention layer splits the problem into a static and dynamic formulation, where the static part is a linear combination of the input data, resulting in the value matrix, $V$, that is used in the attention layer in equation~\ref{eq:attention}.
The observations are used as the query matrix $Q$ and the key matrix $K$ can be interpreted as the adjoint in data assimilation.
The dynamic part adds that information to each member individually by the other members.
A complete attention block uses residual connections that additively update the original member $Z_l$ in its latent space. 
This output is then projected with a final weight matrix $W_o$ back to the output space; hence the attention block can be written as
\begin{equation}
    Z_{l+1} = \sigma\left(Z_l + W_o T(Z_l)\right),
\end{equation}
where $Z$ is the data, $T$ is the attention layer, $W_o$ is the linear projection of the residual output from the attention space to the data domain, and $l$ is the index of the attention block.
 

\begin{figure}[!ht]
    \centering
    \includegraphics[width=\textwidth]{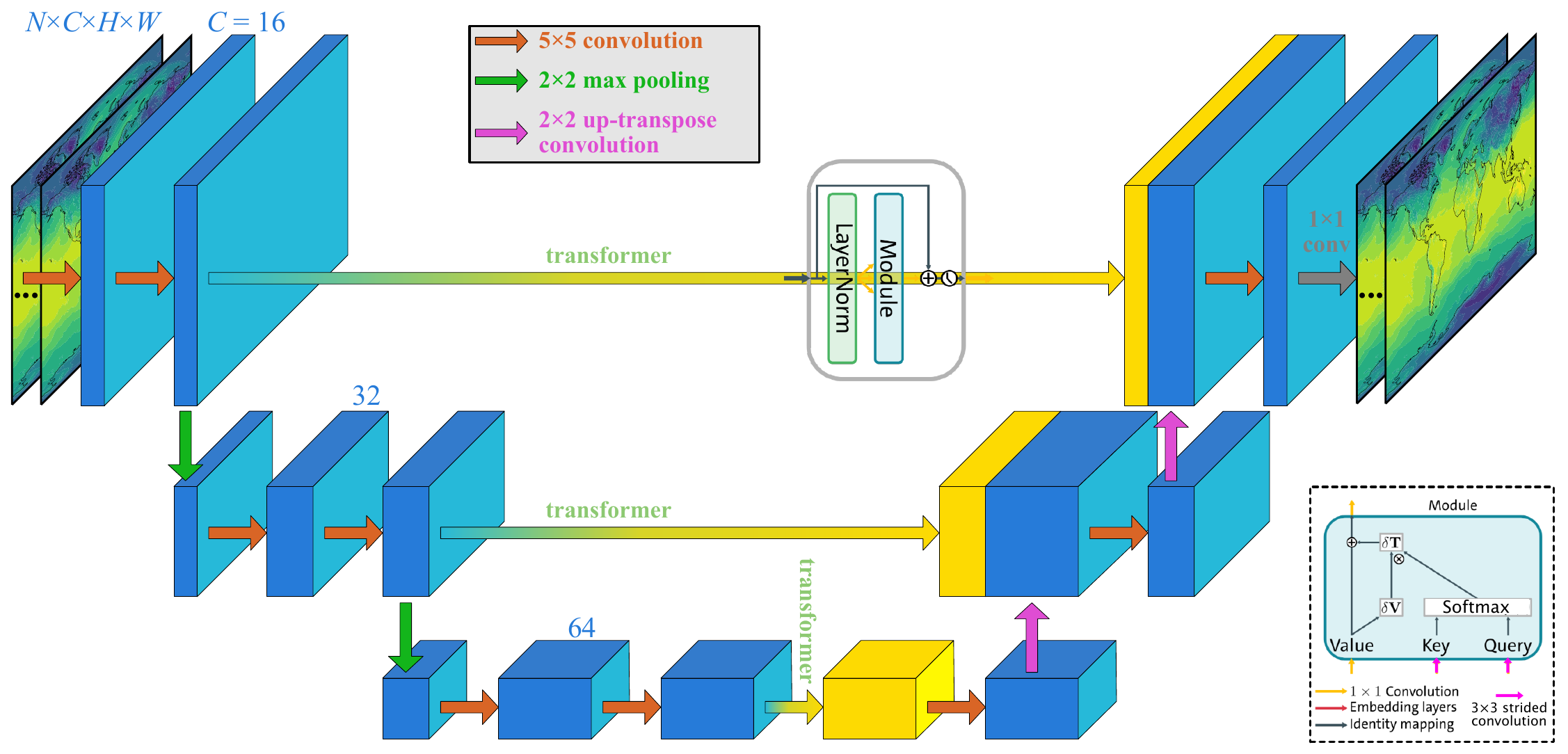}
    \caption{Schematic of the transformer U-net architecture of PoET. The inset on the lower right is adapted from \cite{Finn2021}, used with permission.}
    \label{fig:unet}
\end{figure}

PoET is an adaptation of the ensemble transformer of \cite{Finn2021}. 
The original model was trained on a much smaller dataset, with single input fields and a very coarse resolution of 5.625 degrees in latitude and longitude. 
Our higher-resolution dataset results in a substantial increase in memory cost due to the dot-products across large dimensions $C$, $H$, and $W$. Therefore to manage this we adapt the architecture and implement the transformer within a U-net architecture, shown in Figure\ref{fig:unet}.
At each depth layer in the U-net, following the embedding 2D convolution layers, we add a transformer block\footnote{More than one transformer block can be used, similar to \cite{Finn2021}, but in our experiments this did not result in performance increases.}. 
Within the attention blocks, the convolution layers producing the key and query embeddings use a $3\times3$ convolution with a stride of 3 that serves to further reduce the dimensionality of the dot-product calculation. 
Skip connections after the attention blocks at each level of the U-net allow transformed hidden states to pass through directly to the decoder at multiple spatial resolutions.
Altogether, the PoET implementation reduces the memory footprint of matrix multiplication operations within the transformer and enables the transformers to operate across different spatial scales. 
The layer normalization of the original ensemble transformer still operates at the full resolution of the grid, but unfortunately does not allow the model to be run at a different resolution than that of the training data. 
Experiments omitting the layer normalization or replacing it with another common technique, batch normalization, showed much worse performance. 
This observation cements the layer norm as an integral part of the transformer's ability to correct forecast errors, likely because of its ability to capture local weather effects such as those of topography or land-sea differences.

Finally, some more parameters of the PoET architecture are provided in Appendix~\ref{sec:poetparams} for reference.

\section{Experiments}
\label{sec:experiments}

\subsection{PoET configuration}
\label{sec:poetsettings}

In our experiments, data for lead times every 6 hours from 6 hours up to a maximum of 96 hours are used for training of the 2m temperature model while, for precipitation, we start at 24h to avoid the spin-up. Because the lead time is not explicitly encoded in the model, it is possible to run inference for longer lead times\footnote{While not shown here, PoET remains skillful for longer lead times.}. 

For the post-processing of 2m temperature forecasts, 
we include input features of 2m temperature ($T_2$), temperature at 850 hPa ($T_{850}$), geopotential at 500 hPa ($Z_{500}$), $u$- and $v$-component of winds at 700 hPa ($U_{700}$ and $V_{700}$), and total cloud cover ($TCC$). Additionally, we prescribe orography, a land-sea mask, and the top-of-atmosphere incoming solar radiation (insolation) as additional predictors. Another model using a reduced feature set consisting of only $T_2$, $T_{850}$, and $Z_{500}$, plus the 3 prescribed variables performed only slightly worse than the one trained on the full dataset.

For the post-processing of precipitation forecasts, the input predictors are changed to total precipitation, convective precipitation, convective available potential energy, total cloud cover, total column water, sea-surface temperature, temperature at 850hPa, winds at 700hPa and geopotential at 500hPa.

Apart from the selected predictors, the configuration of PoET is identical for the prediction of 2m temperature and precipitation with two exceptions. Firstly, the normalization of total and convective precipitation is done with a shifted logarithmic transformation\footnote{ $log(x+1)$ with $x$ the precipitation amount normalized into a dimensionless quantity (similar to \citep{lopez2011c})}. This transformation is applied to both the predictor and the predictand total precipitation. The second difference consists of using  the kernel continuous ranked probability score (kCRPS) for precipitation instead of the Gaussian continuous ranked probability score (gCRPS) as a loss function, because the former makes no assumptions on the distribution of the ensemble (the definitions are available in Appendix \ref{sec:crps}). We tested using this formulation for 2m temperature, but observed little difference due to the Gaussian approximation being appropriate.

\subsection{MBM configuration}
\label{sec:mbmsettings}

The MBM parameters are estimated for each grid point and lead time separately. They also vary as a function of the time of the year in order to capture the seasonality of the forecast error. For this purpose, the training dataset differs for each forecast. We define a window centered around the forecast validity date and estimate the parameters using all training data within this time-of-the-year window. The suitable window size is different for the postprocessing of 2m temperature and of precipitation. The window size is set to $\pm$30 days for 2m temperature and to $\pm$60 days for precipitation for all lead times.  
 
As for PoET, a shifted logarithmic transformation of the precipitation data is applied with MBM. Additionally,  in inference mode, spurious precipitation is removed from MBM post-processed precipitation fields and any correction leading to a change in precipitation value greater than 50mm is rejected.

\subsection{Verification process}
\label{sec:verifdef}
We compare PoET, MBM, and raw forecasts in terms of their ability to predict 2m temperature and precipitation up to 4 days in advance. Various aspects of the forecast performance are considered as described below. The results are presented in Section \ref{sec:results} while the formal definitions of the verification metrics can be found in Appendix \ref{sec:skillscore}.

Bias and spread/skill relationships are used to assess the statistical consistency between forecast and verification.
The bias is defined as the average difference between forecast and verification and a reliable forecast has a bias close to zero. The ensemble spread is defined as the standard deviation of the ensemble members with respect to the ensemble mean, while the ensemble mean error is defined as the root mean squared error of the ensemble mean. For a reliable ensemble forecast, the averaged ensemble spread should be close to the averaged ensemble mean error \citep{leutbecher2008ensemble,fortin2014should}. 

The continuous ranked probability score (CRPS) is computed to assess the ensemble as a probabilistic forecast. Forecast performance in a multi-dimensional space is assessed using the energy score (ES), a generalization of the CRPS to the multivariate case \citep{gneiting2007}. ES is applied over the time dimension, computed over 2 consecutive time steps for each pair of time steps separately. Additionally, for precipitation,  probability forecast performance for pre-defined events is assessed with the Brier score \citep[BS][]{brier50}. We consider  2 precipitation events: 6-hourly precipitation exceeding 1mm and 10mm.   

The relative skill of a forecast with respect to a reference forecast is estimated with the help of skill scores. In the following, we compute the continuous ranked probability skill score (CRPSS), the energy skill score (ESS), and the Brier skill score (BSS) using the raw ensemble forecast as a reference. A comparison of PoET and MBM post-processed forecasts is also performed using the latter as a reference.

Scores and reliability metrics are computed at each grid point for all validity times and aggregated in time and/or space for plotting purposes. When aggregating scores in space (over the globe), we apply a weighting proportional to the cosine of the grid point latitude.

\section{Illustrative examples}
\label{sec:examples}

\subsection{PoET in action}
\begin{figure}[!ht]
    \centering
    \includegraphics[width=\textwidth,trim={0cm 5cm 0cm 1.cm},clip]{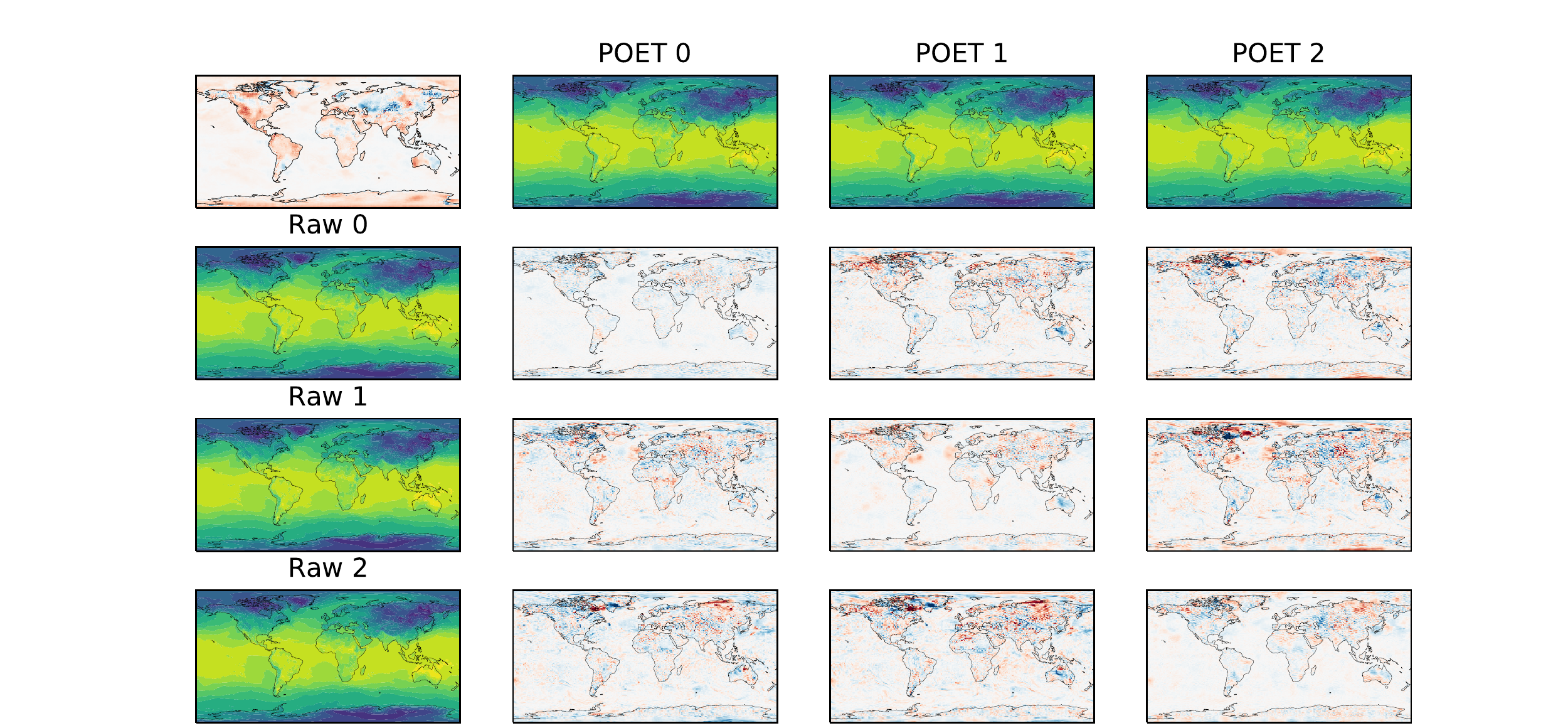}
    \includegraphics[width=\textwidth,trim={0cm 0cm 0cm 14.3cm},clip]{./Global_case_study_cont3.pdf}
    \caption{Relative changes by PoET to a single date of the 2m temperature forecast at day 2, valid on 6 January 2021. The left column shows the first 3 raw ensemble members. The first row shows the first 3 PoET-corrected ensemble members. Other entries show the differences between raw and PoET-corrected ensemble members when the ensemble mean difference has been removed. The ensemble mean difference is plotted in the top left corner panel.}
    \label{fig:Pert-Ensemble-Cross-Comparison}
\end{figure}

Fig.~\ref{fig:Pert-Ensemble-Cross-Comparison} shows the differences between each of the first 3 members of the raw ensemble and the corresponding PoET post-processed forecasts at a lead time of 2 days. The ensemble mean change (the top left panel) mostly shows a global heating, except in Asia which is consistent with the forecast bias discussed in the next section. The top row and left-most column show the PoET forecast and raw forecast, respectively. The remaining entries show the difference between these respective ensemble members once the ensemble mean change is removed. Along the diagonal, we see the change induced by PoET on each member. The off-diagonal entries show the difference between differing raw and PoET ensemble members. The larger amplitude in these off-diagonal plots, compared to the diagonal, indicates consistency between the input and output ensemble members, \textit{i.e.} the ensemble has not been reordered or dramatically shifted by post-processing. A comparison of PoET-corrected forecasts with ERA5 fields in 2 extreme cases is provided below.

\subsection{A 2m temperature case study}
\begin{figure}[!ht]
    \centering
    \includegraphics[width=0.99\textwidth]{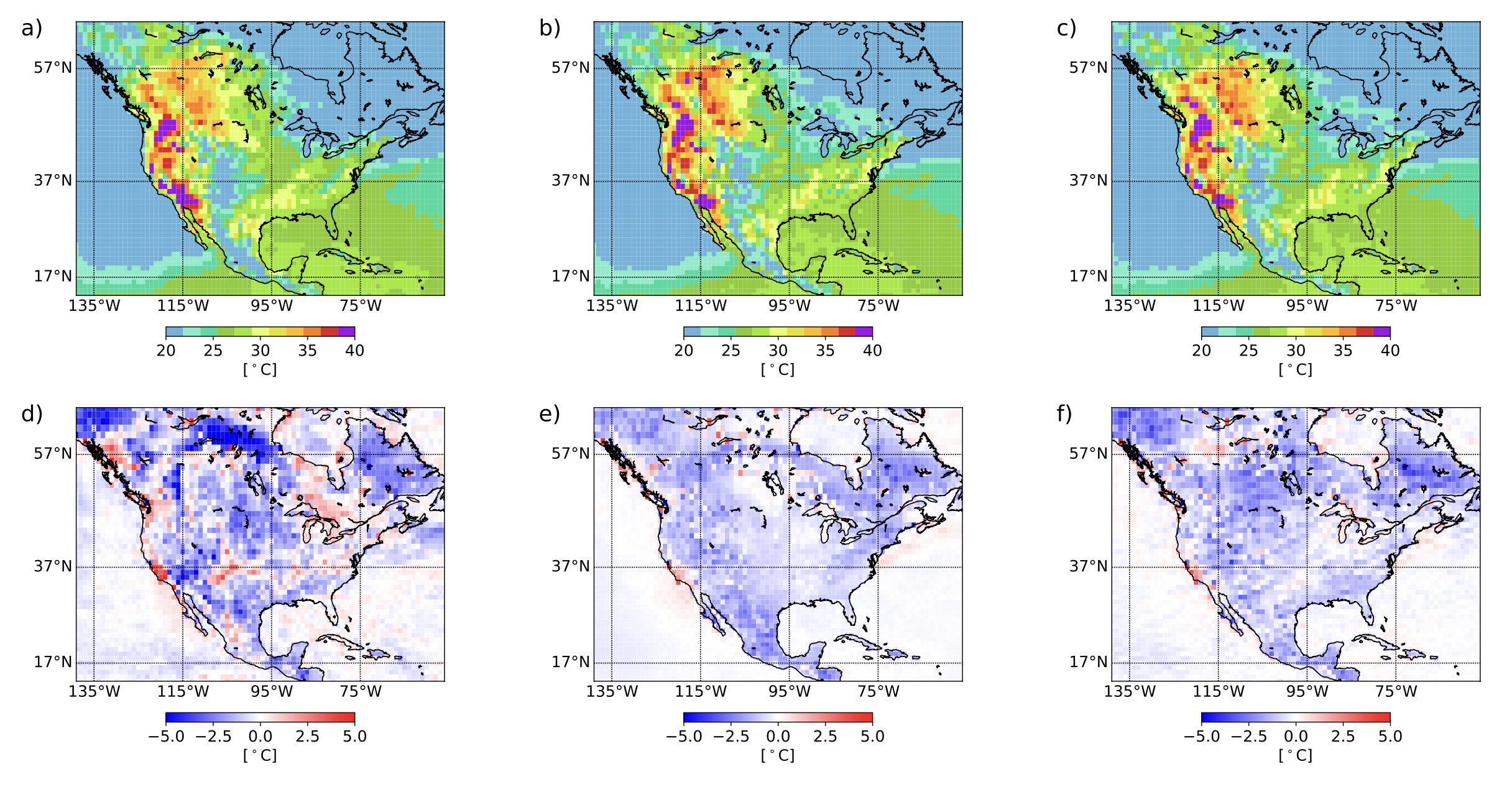}
    \caption{
    3-day averaged maximum temperature between 29 June and 1 July 2021: (a) ERA5, (b) MBM member 21, (c) PoET member 21, (d) difference between the raw forecast and ERA5 ($x^\text{raw} -y $), (e) MBM correction ($x^\text{raw} - x^\text{MBM}$) and (f) PoET correction ($x^\text{raw}- x^\text{PoET}$) made to the raw forecast.  In case of post-processing methods leading to a perfect deterministic forecast, (e) and (f) would match (d). 
    }
    \label{fig:ext2m}
\end{figure}

At the end of June 2021, a heatwave hit the North-western United States and Canada leading to new temperature records and devastating wildfires. The top panels in Fig.~\ref{fig:ext2m} compare the 3-day averaged maximum (00UTC) temperature predictions of MBM and PoET with the corresponding ERA5 reanalysis field. In this example, we average 2m temperature forecasts over lead times 24, 48, and 72h. One randomly selected ensemble member is shown to illustrate post-processing in action on a single forecast. The bottom panels in Fig.~\ref{fig:ext2m} show the difference between ERA5 and the raw forecast as well as the corrections applied to the forecast with post-processing. We check whether post-processing compensates for errors in the raw forecast, that is if Figs \ref{fig:ext2m}(e) and \ref{fig:ext2m}(f) match Fig. \ref{fig:ext2m}(d). Overall, there is a good correspondence between the raw forecast error and the post-processing corrections for both MBM and PoET. For example, we note the correction of the cold bias over the continent. However, there is some spottiness visible in the PoET correction (Fig. \ref{fig:ext2m}f). Moreover, in both Fig. \ref{fig:ext2m} (e) and (f), as expected, fine details in the error pattern are not accurately captured, due to factors such as the limited predictability of this extreme event.

\subsection{A total precipitation case study}
\label{sec:tpex}
\begin{figure}[!ht]
    \centering
    \includegraphics[width=0.99\textwidth]{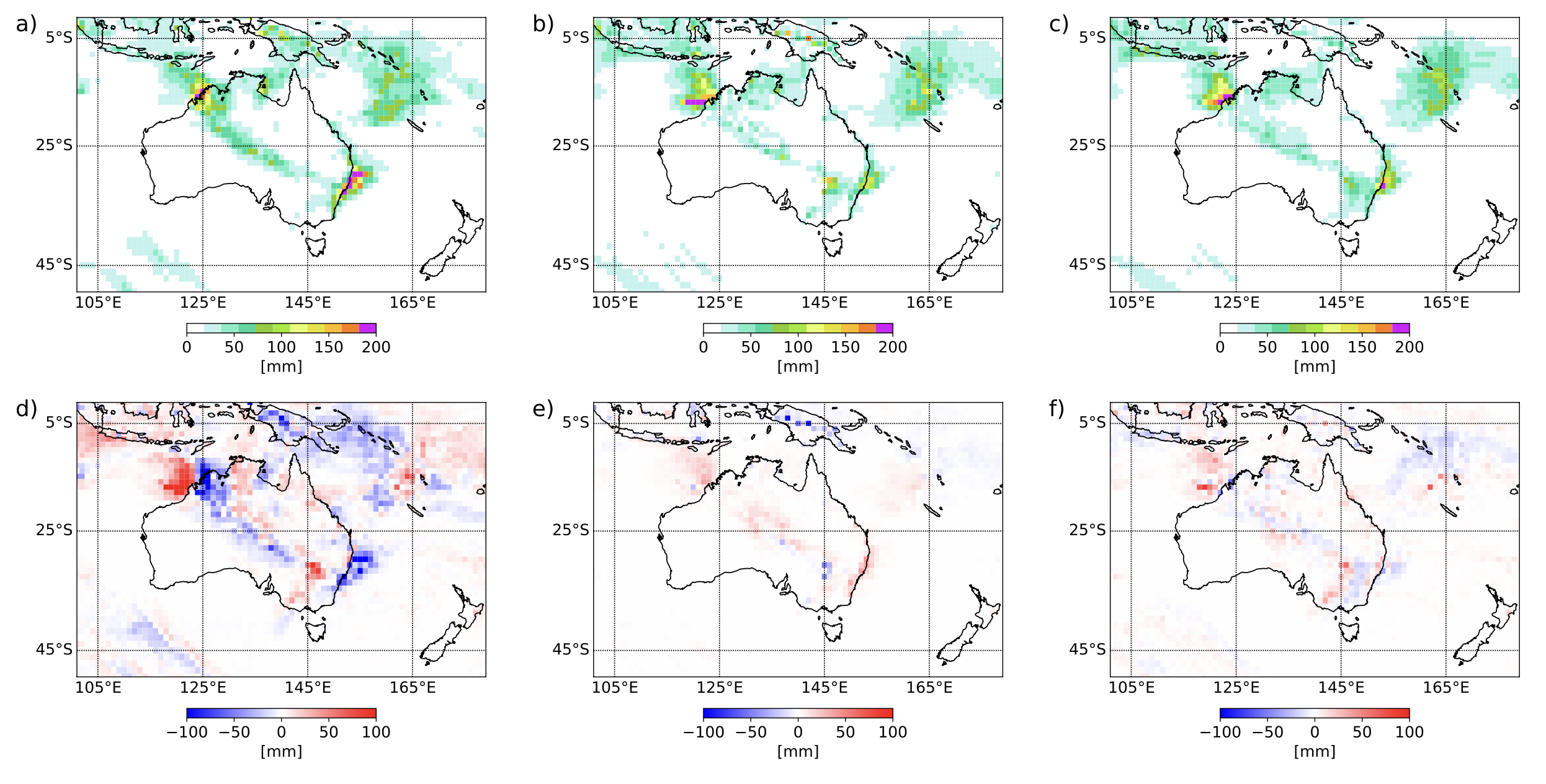}
    \caption{ Same as Fig.~\ref{fig:ext2m} but for 3-day accumulation precipitation between 18 March and 21 March 2021.
    }
    \label{fig:exprecip}
\end{figure}

In March 2021, Australia was affected by extreme rainfall. Sustained heavy rain led to flooding in the Eastern part of the country and large precipitation amounts were observed on the Northern coast too. The top panels in Fig.~\ref{fig:exprecip} compare 3-day precipitation scenarios from MBM and PoET with the corresponding ERA5 precipitation field. The 3-day accumulated precipitation scenarios are derived from the post-processed ensemble of 6h accumulated precipitation forecasts that are consistent scenarios in space and time. Here again, we randomly select one member for illustration purposes. The bottom panels in Fig. \ref{fig:exprecip} show the difference between ERA5 and the raw precipitation forecast along the corresponding post-processing corrections. 
As in the 2m temperature example, we check whether post-processing compensates for raw forecast errors, \textit{i.e.} if Figs \ref{fig:exprecip}(e) and \ref{fig:exprecip}(f) match Fig. \ref{fig:exprecip}(d). The MBM correction only has some areas of consistency with the actual error while the PoET correction tends to partially compensate for the raw forecast error along the North and West coast, both over land and over the sea.

\section{Verification results}
\label{sec:results}

\subsection{2m temperature results} 

\begin{figure}[!ht]
    \centering
    \includegraphics[width=0.95\textwidth]{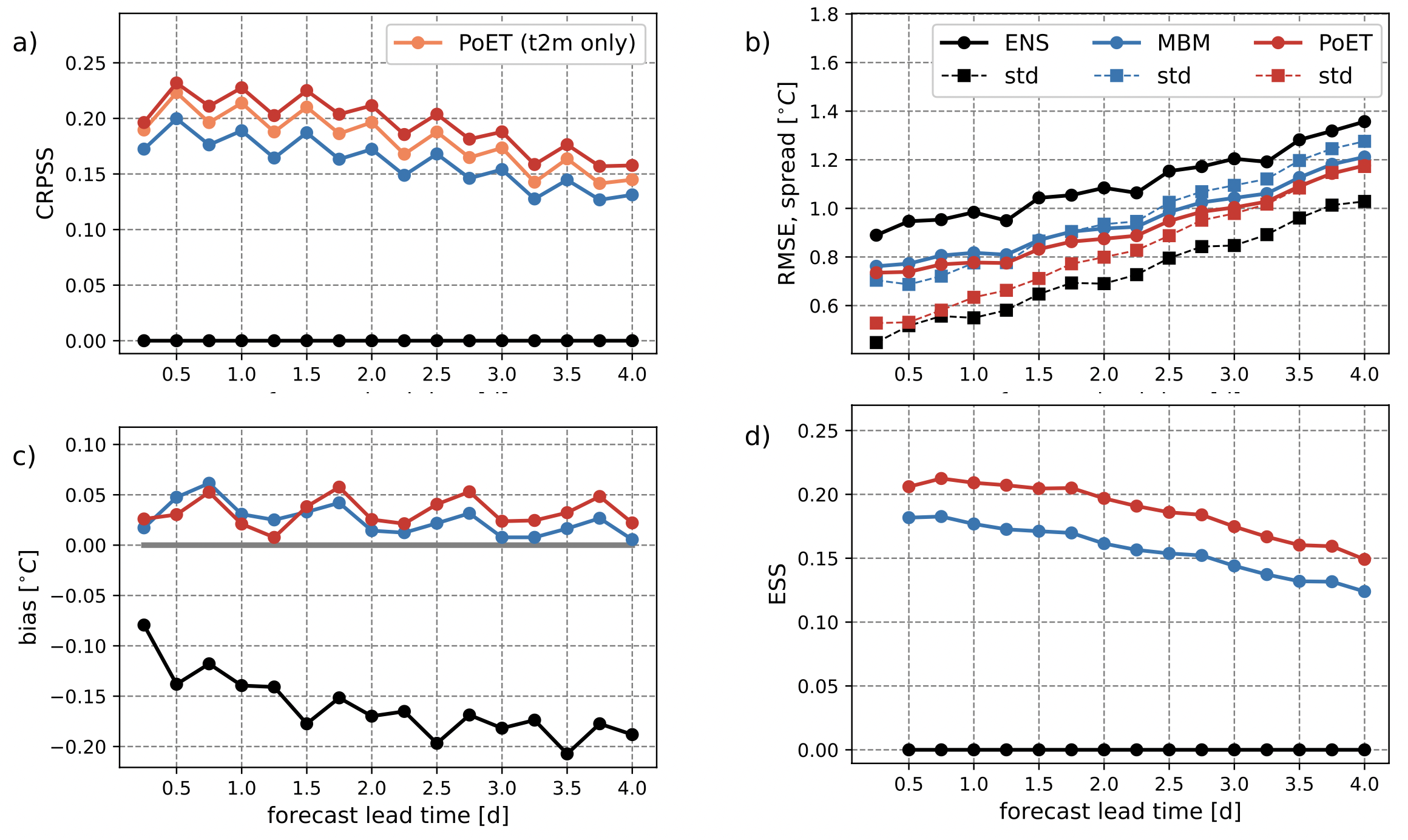}    
     \caption{(a)  CRPSS (the higher, the better), (b) spread and skill, (c) bias (optimal value zero), and (d) ESS (the higher, the better) of 2m temperature for 3 ensemble forecasts: raw ensemble, MBM, and PoET. In (a), we also show results for PoET using only 2m temperature (t2m) as a predictor. }
    \label{fig:spread-skill-t2m}
\end{figure}

In order to compare and assess the performance of MBM and PoET, we apply the verification metrics defined in Section \ref{sec:verifdef} (\textit{i.e.} the CRPSS, spread and error, bias, and ESS) to the post-processed forecasts. 
The results are first aggregated over the globe as a function of the forecast lead time. In Fig.~\ref{fig:spread-skill-t2m}, we see that both methods considerably improve the raw ensemble skill with similar results in terms of CRPSS and ESS. PoET generates more skillful forecasts than MBM, but both methods are able to improve by $\sim~20$\% the raw forecast throughout the assessed lead times. Both methods also have a similar ability to reduce the bias. 

An additional experiment is run to disentangle the benefit of using the new ML-based method from the benefit of having more than one predictor as an input variable. Post-processed forecasts are generated with PoET using 2m temperature only as a predictor similarly as when running MBM and the results in terms of CRPSS are shown in Fig.~\ref{fig:spread-skill-t2m}(a). The use of the PoET machinery for post-processing seems to contribute to around two thirds of the improvement over the benchmark method while the remaining improvement can be attributed to extra information available in the additional predictors. This result appears consistent over lead times. 

The MBM approach seems better at maintaining a spread-error parity, with PoET struggling at early lead times. Spread/skill diagrams, showing the error as a function of spread categories, reveal that aggregated scores must be interpreted carefully (see Appendix \ref{sec:relplots}). Indeed, the uncertainty of PoET-corrected forecasts appears to reflect the potential forecast error more accurately than the MBM-corrected ones. Because compensating effects can be at play when averaging over all cases, we also compute bias and spread error ratio at each grid point separately before averaging absolute terms over the verification period (see Appendix \ref{sec:mabsplots}). This approach reveals that PoET calibration underperformance compared with MBM is moderate and limited to the first lead times of the forecast. This result suggests a geographical disparity of the post-processing impact that is now further explored with maps of scores.

\begin{figure}[!ht]
    \centering
\includegraphics[width=0.99\textwidth]{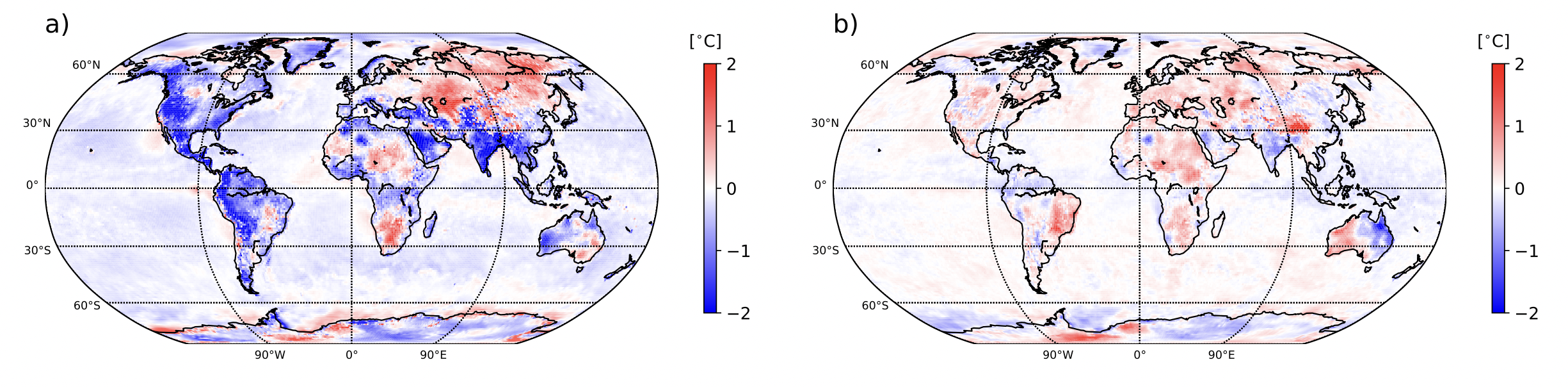}   
\caption{
    Bias of  2m-temperature forecasts at lead time day 4 for (a) the raw ensemble and (b) PoET. An optimal forecast has a bias close to 0.
    }
    \label{fig:bias2t}
\end{figure}

Fig.~\ref{fig:bias2t} shows maps of bias for the raw data and the PoET-corrected forecasts.  We focus on lead time day 4 and the results are aggregated at each grid point over all verification days.  We clearly see a general decrease in the bias with almost no remaining bias over the ocean. The remaining pockets of (generally positive) bias after post-processing are mostly found over land where the amplitude of the raw forecast bias is larger. A change of sign in the bias is interpreted as an indication that the general circulation patterns over the training is not representative of the ones over the verification period. The broad structure of the bias is similar for MBM and for other lead times (not shown).

\begin{figure}[!ht]
    \centering
    \includegraphics[width=0.99\textwidth]{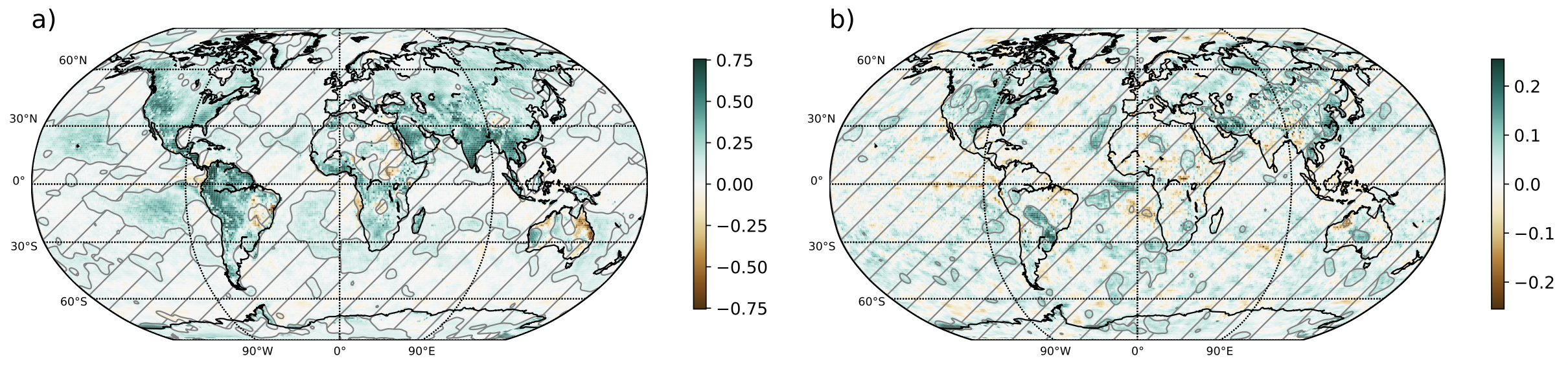}   
    \caption{
    CRPSS of 2m-temperature PoET forecast with respect to (a) the raw ensemble and (b) MBM. Positive values indicate a skill improvement with PoET. Note the difference in scale between the 2 plots.
    }
    \label{fig:crpss2t}
\end{figure}

Fig.~\ref{fig:crpss2t} shows the gain in skill with PoET for the same lead time as in Fig.~\ref{fig:bias2t}. CRPSS is computed using the raw ensemble as a baseline in Fig.~\ref{fig:crpss2t}(a) and MBM as a baseline in Fig.~\ref{fig:crpss2t}(b). 
Fig.~\ref{fig:crpss2t}(a) shows a widespread positive impact of PoET on the raw forecast skill with a larger gain over land where the raw forecast bias is generally more pronounced. A detrimental effect of post-processing is observed in some regions (\textit{e.g.} in South America, Africa, and Australia). These regions of negative skill score are also the ones where a bias is still present after post-processing as shown in Fig.~\ref{fig:bias2t}(b).

In Fig.~\ref{fig:crpss2t}(b), there are very few areas where the CRPSS is less than zero, \textit{i.e.} areas where MBM forecasts have more skill than PoET. Improvements through PoET are fairly consistent across the globe, with no regions where there are larger gains due to the neural network approach. Indeed, there is a  strong agreement between the locations where MBM and PoET add value to the raw ensemble (not shown). Given that MBM learns a climatological correction for each grid point this suggests PoET has mostly reproduced this climatological local correction.

\subsection{Total precipitation results}
Post-processing of precipitation forecasts is a more challenging task because of the form of the underlying forecast probability distribution with a point mass at 0 (the no-precipitation probability) and a skewness capturing the more extreme events. Post-processing with MBM and PoET is tested with small changes to the configuration used for the post-processing of 2m temperature forecast (see Section \ref{sec:experiments}), and a similar set of plots is examined to assess the corrected forecast performance. 

\begin{figure}[!ht]
    \centering
    \includegraphics[width=0.92\textwidth]{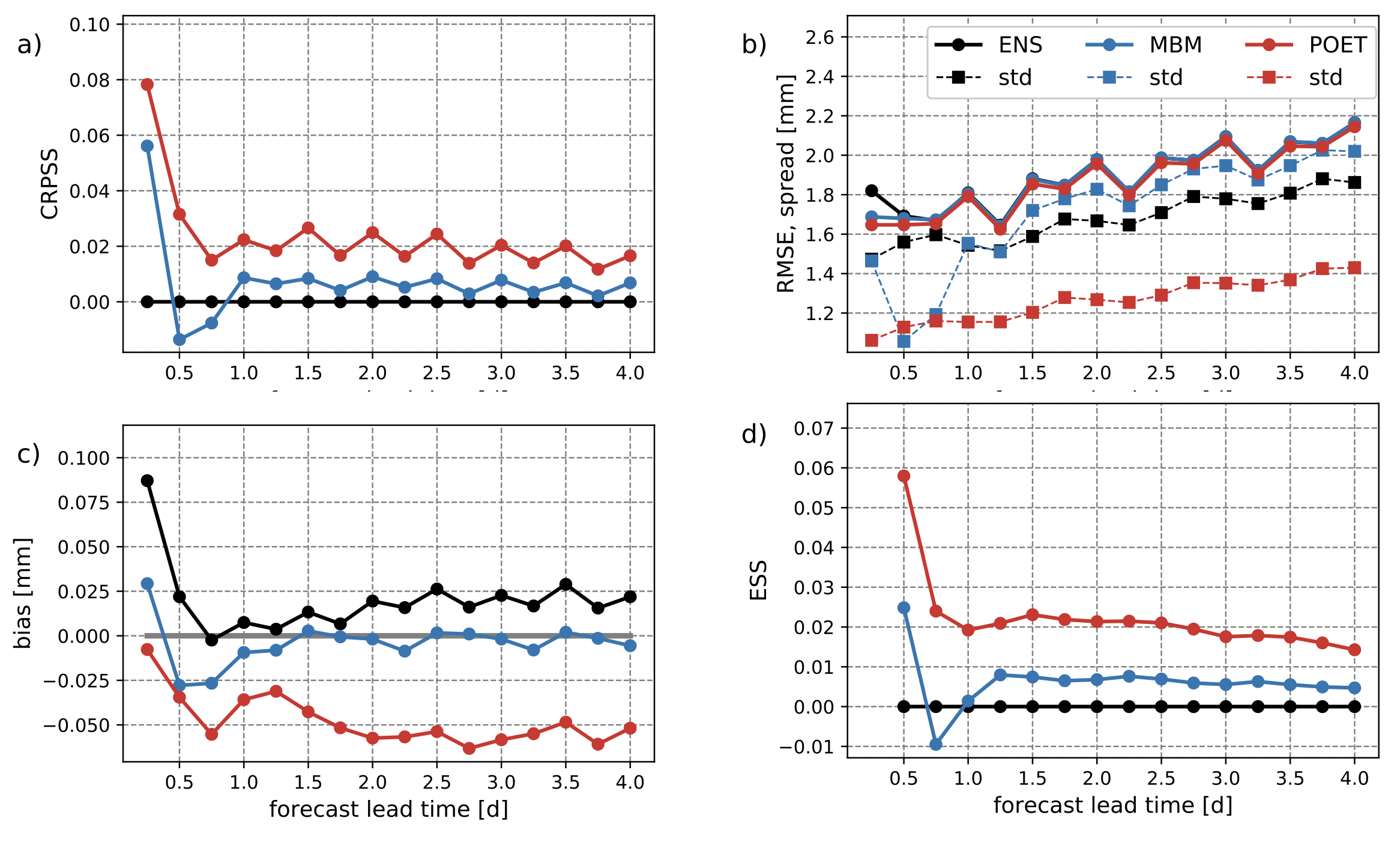}
    \caption{Same as Fig.~\ref{fig:spread-skill-t2m} but for total precipitation.}
    \label{fig:spread-skill-tp}
\end{figure}

Fig. \ref{fig:spread-skill-tp} shows verification metrics aggregated globally as a function of the lead time. In contrast to 2m temperature results, the added benefit of either postprocessing approach is limited.  With PoET, the skill improvement is approximately 2\% in terms of CRPSS and ESS for the first several days of forecasting. With MBM, the skill improvement is  $\sim$1\% for most lead times. The gain in skill originates from improved performance in forecasting lower-intensity rather than higher-intensity events. Indeed, BSS computed for 2 precipitation exceeding thresholds, 1mm and 10mm, shows a larger skill score for the former (see Fig. \ref{fig:brierscore} in Appendix).

One explanation for the limited gain in skill with post-processing is that the raw forecast is already well-calibrated (see also Fig.~\ref{fig:reliability}(c) and (d) in the Appendix). 
Also, PoET improves the averaged performance in Figs \ref{fig:spread-skill-tp}(a) and (d) but seems to degrade both the bias and the spread-error ratio in Figs \ref{fig:spread-skill-tp}(b) and (c).
This apparent paradox is explained by the large variations in forecast performance over the globe. A look at the mean absolute bias and mean absolute spread bias confirms that bias and spread/skill relationship are overall significantly improved with PoET when assessed locally (see Fig.~\ref{fig:new_tp} in Appendix). Similarly, the erratic spread correction with MBM at a shorter lead time is not visible in Fig.~\ref{fig:new_tp}(b) suggesting an averaging artifact. Fig.~\ref{fig:new_tp}(b) also reveals that a point-by-point application of MBM does not seem appropriate to correct spread deficiencies at longer lead times.

\begin{figure}[!ht]
    \centering
\includegraphics[width=0.99\textwidth]{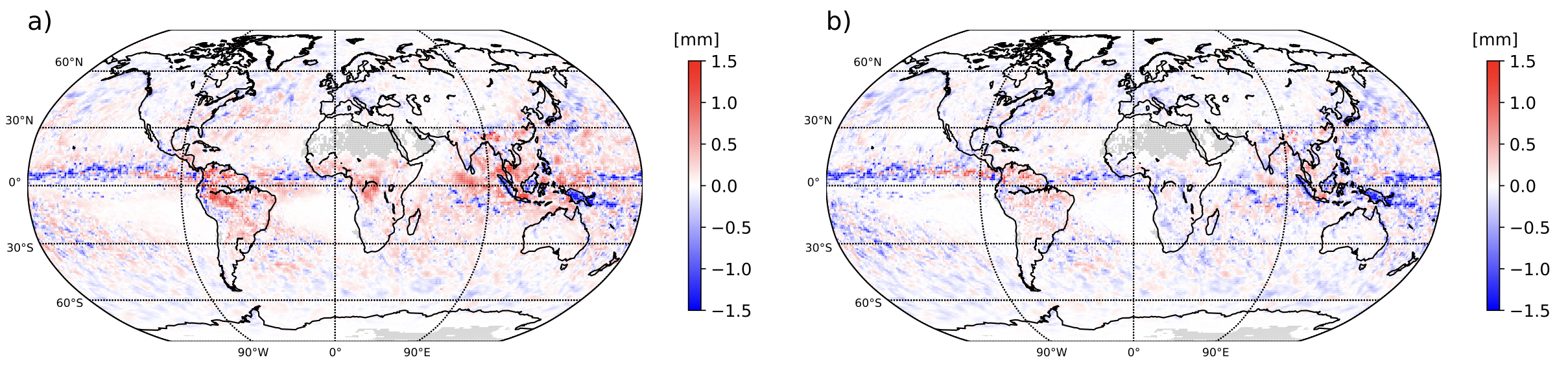}   
\caption{
    Same as Fig.~\ref{fig:bias2t} but for total precipitation. A bias close to 0 is optimal. Regions with annual precipitation lower than 0.1mm are masked in grey.
    }
    \label{fig:biastp}
\end{figure}

Fig.~\ref{fig:biastp} provides another perspective on the bias by presenting maps of averaged values over all verification days. Here, the focus is on lead time day 4. The precipitation bias is reduced over land and the maritime continent. However,  the bias in the tropical Pacific and Atlantic remains unchanged after post-processing. We note that whilst positive biases are generally well corrected, negative biases are not.

\begin{figure}[!ht]
    \centering
    \includegraphics[width=0.99\textwidth]{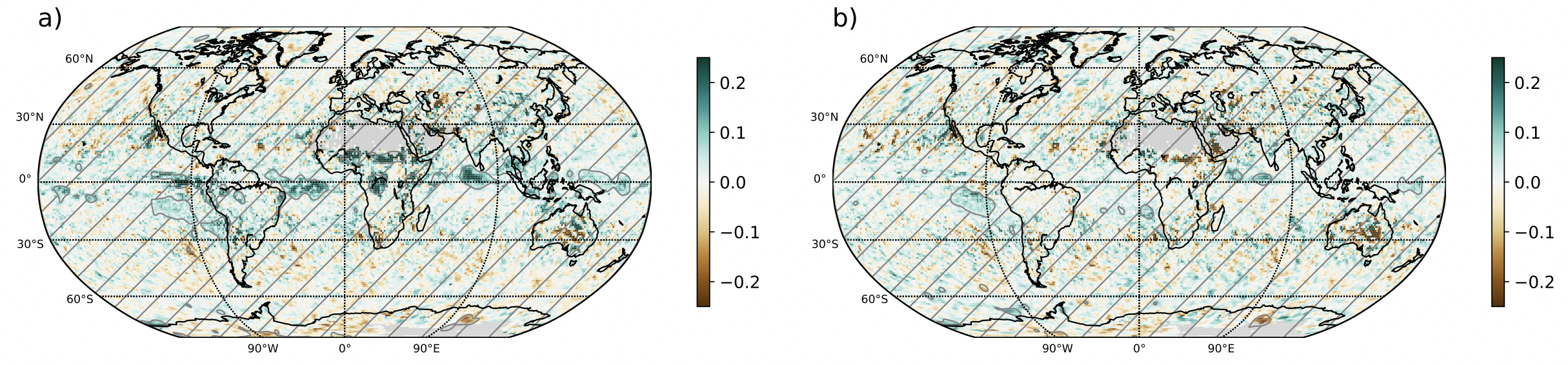}
    \caption{
    Same as Fig.~\ref{fig:crpss2t} but for total precipitation. Positive values indicate a gain in skill with PoET. Regions with annual precipitation lower than 0.1mm are masked in grey.
    }
    \label{fig:crpsstp}
\end{figure}

Finally, Figs \ref{fig:crpsstp}(a) and \ref{fig:crpsstp}(b) show maps of CRPSS at day 4 for PoET using the raw forecast and MBM as a reference, respectively. PoET improves precipitation ensemble forecasts mainly over the tropics.  Very localized degradation of the skill with respect to the raw forecast could be due to a too short training sample. The benefit of using PoET rather than MBM appears predominantly in the tropics but local positive skill scores are rather scattered. Alternate areas of positive and negative skill scores over the sea in the extra-tropics suggest that the 2 approaches are complementary. 

\subsection{\textit{ENS-10} comparison}
\label{sec:ens10}
During the preparation of the manuscript at hand, \cite{ens10} produced a benchmark dataset for the postprocessing of ensemble weather forecasts (referred to as \textit{ENS-10} in the following). This framework is exploited to further test the capability of PoET and compare its performance with state-of-the-art ML post-processing techniques.

\textit{ENS-10} dataset is similar to the one focused on here, originating from the same model reforecast framework, but with several differences. The reforecast dataset is constructed in 2018, the spatial resolution of the data is provided on a 0.5$^\circ$ grid, and the evaluation set comprises the last 2 years of the reforecast, meaning that the IFS configuration is identical between training and testing, also in terms of ensemble size. Note that the verification differs as a non-latitude-weighted CRPS is used as a performance metric. 

To contribute to the benchmarking efforts, we train our model using the \textit{ENS-10} dataset and evaluate following the same methodology. We reduce the data volume by only training each model on a chosen subset of the total ENS10 variables. For Z500, we utilize all ENS10 variables on this pressure level and use an equivalent approach for T850. For T2m, our model predictors are 500hPa, 850hPa alongside single level variables 2m temperature, skin temperature, sea surface temperature, mean sea level pressure, total cloud cover and 10m zonal and meridional wind. Also, the gridded data contains 720 points located at each pole that are unconstrained by the latitude-weighted training with PoET but contribute to the non-latitude-weighted evaluation. Therefore, we do not use PoET to correct these points but instead use the uncorrected raw forecast (the evaluation still includes these points to mirror \textit{ENS-10} evaluation). 

In terms of model architecture, we make a small change to the PoET model to incorporate the higher spatial resolution. We increase the depth of the UNet structure by 1, putting a Transformer block at its fourth level.

Table \ref{tab:benchmarks_full} contrasts the PoET scores with the raw forecast and the best benchmarks from \cite{ens10}. For almost all configurations, the PoET approach leads to significant improvement over the ENS-10 benchmarks. In particular, for 2m temperature, the CRPSS with PoET is considerably better than the results with the previous baselines.  The improvement of $\sim20$\% is similar to the gain measured with our dataset.
For Z500 with a 5-member ensemble, the model improves the raw output but fails to beat the LeNet approach. We found no explanation for the limited success with this variable and ensemble member configuration. Similar results are obtained with 
the extreme event weighted continuous ranked probability score (EECRPS) introduced in Eq.~(2) in \cite{ens10}.

\begin{table*}[t]
  \centering
  \includegraphics[width=0.92\textwidth]{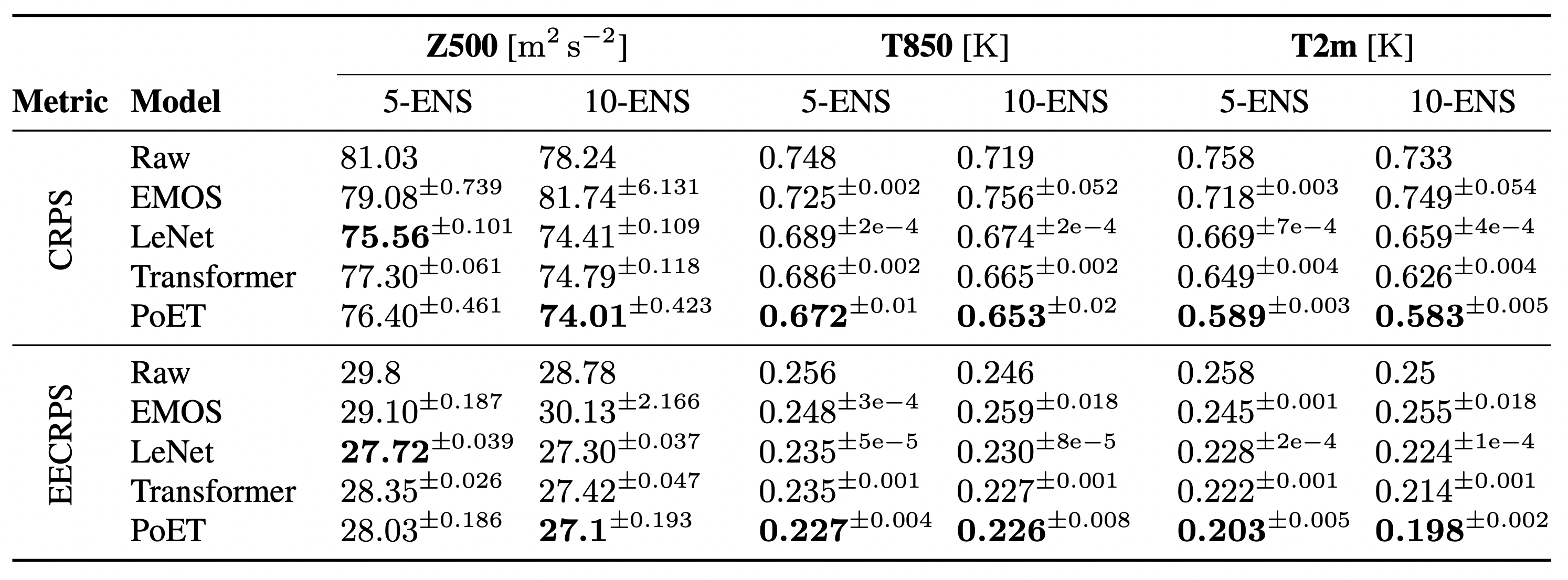}
  \caption{Global non-latitude-weighted mean CRPS and EECRPS on the ENS-10 test set (2016--2017) for baseline models with five (5-ENS) and ten (10-ENS) ensemble members. The first 5 members are used for the 5-ENS evaluation.}
  \label{tab:benchmarks_full}
\end{table*}

\section{Conclusion}
\label{sec:conclusion}
This work shows how to efficiently transform an ensemble forecast into a calibrated ensemble forecast \textit{i.e.} a set of physically consistent scenarios with improved statistical properties.  We compare two methods: one machine-learning method based on self-attention transformers (PoET) and one statistical method used as a benchmark (MBM). For both methods, each member is calibrated separately but with the aim of optimizing the ensemble properties as a whole. As a result, the post-processed ensemble has the spatial, temporal, and inter-variable coherence necessary to enable any downstream application. Also, both tested methods can be trained on a smaller reforecast dataset (here with 11 members) to effectively calibrate a much larger operation ensemble  (here with 51 members), preserving inter-member calibration. 

Ensemble post-processing is successfully applied to global gridded forecasts of 2m temperature and precipitation, using ERA5 reanalysis as the ground truth. Our results show that both MBM and PoET can significantly improve the skill of the operational IFS ensemble. This improvement is achieved through a better calibration of the ensemble, both in terms of bias and spread-skill relationship. We note that PoET is better at the headline scores (CRPS, ES) but with some areas where MBM can locally outperform PoET. This latter point suggests that a combination of the two approaches could lead to further improvement of the forecast skill. Also, our case-study examples illustrate the ability of post-processing to improve existing ensemble members.

The post-processing gain is smaller for precipitation than for 2m temperature. Indeed, the skill improvement of precipitation forecasts is relatively small in this application. This result contrasts with results obtained with down-scaling approaches where accounting for representativeness uncertainty can have a major impact on scores \citep{zbb2020}. In further work, we will consider how PoET could be applied to un-gridded observations, which would require architectural changes.

Direct applications of the methodologies developed here include post-processing for forecast verification and intercomparison purposes. For example, bias correction can be applied to better understand changes in CRPS results with new IFS model versions \citep{leutbecher2021}. Also,  post-processing would be a necessary step for a fair comparison of NWP forecasts with statistically optimized (data-driven) ones in forecasting competition frameworks \citep[see for example][]{rasp2020}. Finally, the proposed methods could be trivially adapted to a higher-resolution version of the truth, which could pave the way to ensemble post-processing of global gridded data for operational forecasting.

\nocite{*}
\section*{Acknowledgements}
The authors thank Tobias Finn for many interesting discussions and the original idea of using transformer techniques in the ensemble dimension. We also thank 2 anonymous reviewers for their valuable suggestions to improve the quality of this paper

Peter Dueben, Matthew Chantry and Jesper Dramsch gratefully acknowledge funding from the MAELSTROM EuroHPC-JU project (JU) under No 955513. The JU receives support from the European Union’s Horizon research and innovation program and United Kingdom, Germany, Italy, Luxembourg, Switzerland, and Norway. Peter Dueben gratefully acknowledges funding from the ESiWACE project funded under Horizon 2020 No. 823988. Mariana Clare gratefully acknowledges funding by the European Union under the Destination Earth initiative. Finally, all authors acknowledge the use of compute resources from both the European Weather Cloud and Microsoft Azure.

\section*{Data and Code availability}
It is currently difficult for the authors to share the data as the hardware used is currently being replaced. The authors will, however, make the data available for download when the paper is eventually published.

PoET source code will be made available soon, the MBM parameter estimation relies on the Climdyn/pythie package  \citep[][]{demaeyer2022}.

\bibliographystyle{ametsocV6}

\vspace{1.5cm}
\section*{Appendix}
\vspace{0.5cm}

\subsection{MBM parameters optimization}
\label{sec:mbmopt}
The MBM parameters $\alpha$ and $\beta$ are derived by resolving an ordinary least-square regression. They are computed using 
\begin{equation}
\alpha = \langle {y}\rangle - \langle \beta\overline{x}\rangle,
\end{equation}
with $y$ the verification and where $\langle \cdot \rangle$ is an averaging operator applied over the training sample,
\begin{equation}
\beta= \rho_{y\overline{x}}\frac{\sigma_{y}}{\sigma_{\overline{x}}},
\end{equation}
with $\rho_{y\overline{x}}$ the correlation between  verification and ensemble mean, $\sigma_{y}$ and  $\sigma_{\overline{x}}$ the variance of the verification and of the ensemble mean, respectively.
Imposing the reliability constraints and using maximum likelihood estimators, we have
\begin{equation}
\gamma^2 = \frac{ \sigma^2_{y}}  {\langle \sigma_\epsilon^2 \rangle} ( 1- \rho_{y\overline{x}}^2 )
\end{equation}
with $\sigma_{\epsilon}$ the ensemble spread (ensemble standard deviation). 

\subsection{Additional PoET model parameters}
\label{sec:poetparams}
Table~\ref{tab:poet} shows some additional hyperparameters used for the PoET model. Minimal searching was performed to obtain these parameters as no significant improvements were observed.

\begin{table}[t]
  \centering
  \small
  \begin{tabular}{@{}c@{\hspace{\tabcolsep}}ll@{\hspace{\tabcolsep}}ll@{\hspace{\tabcolsep}}ll@{\hspace{\tabcolsep}}l@{}}
  \toprule
    Parameter & Value \\
  \midrule
    Attention blocks per transformer module & 1 \\
    Number of transformer modules & 3 \\
    Attention heads & 8 \\
    Activations & ReLU \\
    Optimizer & Adam \\
    Learning rate & $1\times 10^{-3}$ \\
    Early stopping & 15 epochs \\
    Learning rate schedule & Plateau (15 epochs) \\
  \bottomrule
  \\
  \end{tabular}
  \caption{Additional hyperparameters for PoET.}
  \label{tab:poet}
\end{table}

\subsection{Scores definition}

\subsubsection{Continuous ranked probability score (CRPS)}
\label{sec:crps}

 Following \cite{gneiting2007}, the kernel CRPS is defined as
\begin{equation}
kCRPS~\left(\mathbf{x},y\right) = \frac{1}{M} \sum_{i=1}^M | x_i - y | - \frac{1}{2M^2} \sum_{i=1}^M \sum_{j=1}^M | x_i - x_j |,
\end{equation}
where $\mathbf{x}$ is an ensemble of size $M$ predictions, and $y$ is the outcome.

\cite{gneitingCalibratedProbabilisticForecasting2005} suggested a closed-form expression for the CRPS of a Gaussian distribution with mean $\mu$ and variance $\sigma^2$ defined as
\begin{equation}
gCRPS (\mathcal{N}(\mu,\sigma^2),y ) =  \tfrac{\sigma} {\sqrt{\pi}} \left( \sqrt{\pi} \tfrac{y-\mu} {\sigma} \text{erf} \left(\tfrac{y-\mu} {\sqrt{2}\sigma} \right) + \sqrt{2} \text{exp} \left(-\tfrac{(y-\mu)^2} {2\sigma^2} \right) -1 \right)
\end{equation}
with 
\begin{equation}
\text{erf} = \tfrac{2}{\sqrt{\pi}}\int_{0}^{x}exp(-x^2).
\end{equation}

\subsubsection{Energy score (ES)}
\label{sec:energyscore}
ES is a generalization of the CRPS to the multivariate case \citep{gneiting2007}. It is defined as 
\begin{equation}
    \text{ES}(\mathbf{X},\mathbf{y})= \frac{1}{M}\sum_{j=1}^M  \left\Vert \text{X}^j - \text{y} \right\Vert  - \frac{1}{2M^2}\sum_{i=1}^M\sum_{j=1}^M  \left\Vert \text{X}^i - \text{X}^j \right\Vert
\end{equation}
where $\mathbf{X}$ and $\mathbf{y}$ are the ensemble forecast and verification in the multivariate space, respectively. 

\subsubsection{Skill scores}
\label{sec:skillscore}
Skill scores measure the relative performance of a given forecast $f$ with respect to a \textit{reference} forecast $g$. For a given score, its skill score $SS$ is computed as
\begin{equation}
  SS = 1 - \frac{\overline{S}_f}{\overline{S}_g}
\end{equation}
with $\overline{S}_f$ and $\overline{S}_g$ the averaged score $S$ for forecasts $f$ and $g$, respectively. 

\subsection{Additional plots}
\label{sec:supp}
\subsubsection{Reliability diagrams}
\label{sec:relplots}
\begin{figure}
    \centering
    \includegraphics[width=0.98\textwidth]{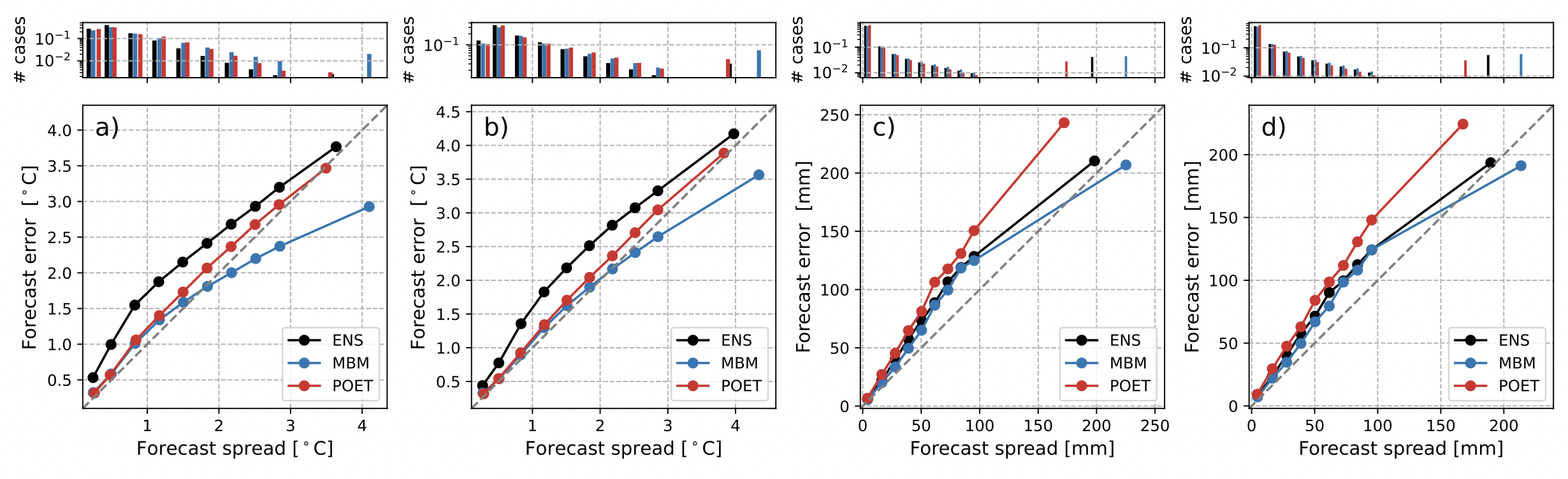}
    \caption{Reliability plots showing the spread/error relationship for 2m temperature forecasts at lead time day 2 (a) and day 4 (b), and for precipitation forecasts at lead time day 2 (c) and day 4 (d).  For each plot, a histogram shows the number of cases in each forecast uncertainty category. }
    \label{fig:reliability}
\end{figure}

For a reliable ensemble forecast, we expect consistency between ensemble spread and ensemble mean error. This can be checked with reliability plots as shown in Fig.~\ref{fig:reliability}. For a given forecast spread category (on the x-axis), we check the consistency of the corresponding forecast error (RMSE on the y-axis). Perfect reliability is indicated with a dashed diagonal line.

\subsubsection{Mean absolute bias and mean absolute spread bias}
\label{sec:mabsplots}
A bias computed over a heterogeneous sample can be misleading because of the compensating effects at play. As a complementary verification metric to the averaged bias, we compute a mean absolute bias as follows: 
\begin{equation}
  \Biggl  \langle \Bigl\vert \langle \overline{x} - y \rangle_{time} \Bigl\vert \Biggl \rangle_{space}
\end{equation}
where $\langle.\rangle_{time}$ and $\langle.\rangle_{space}$ are the averaging operator over the verification period and verification domain, respectively. The bias is computed spatially (at each grid point) and then averaged over the verification period in mean absolute terms.  

Similarly, a comparison of the ensemble spread with the ensemble mean error might not be meaningful if spread and error are averaged over a heterogeneous sample. Here, we suggest computing the ratio between spread and error at each grid point separately before centering around 0 and averaging over time in absolute terms. Formally, the mean absolute spread bias is computed as follows:
\begin{equation}
  \Biggl  \langle \Bigl\vert \frac{\sqrt{\langle \frac{M+1}{M-1}\sum_i^M( x_i - \overline{x})^2 \rangle_{time}}} {\sqrt{\langle ( y - \overline{x})^2 \rangle_{time}}} -1 \Bigl\vert \Biggl \rangle_{space}.
\end{equation}
where the factor $\frac{M+1}{M-1}$ accounts for the limited ensemble size.

\begin{figure}[ht]
    \centering
    \includegraphics[width=0.93\textwidth]{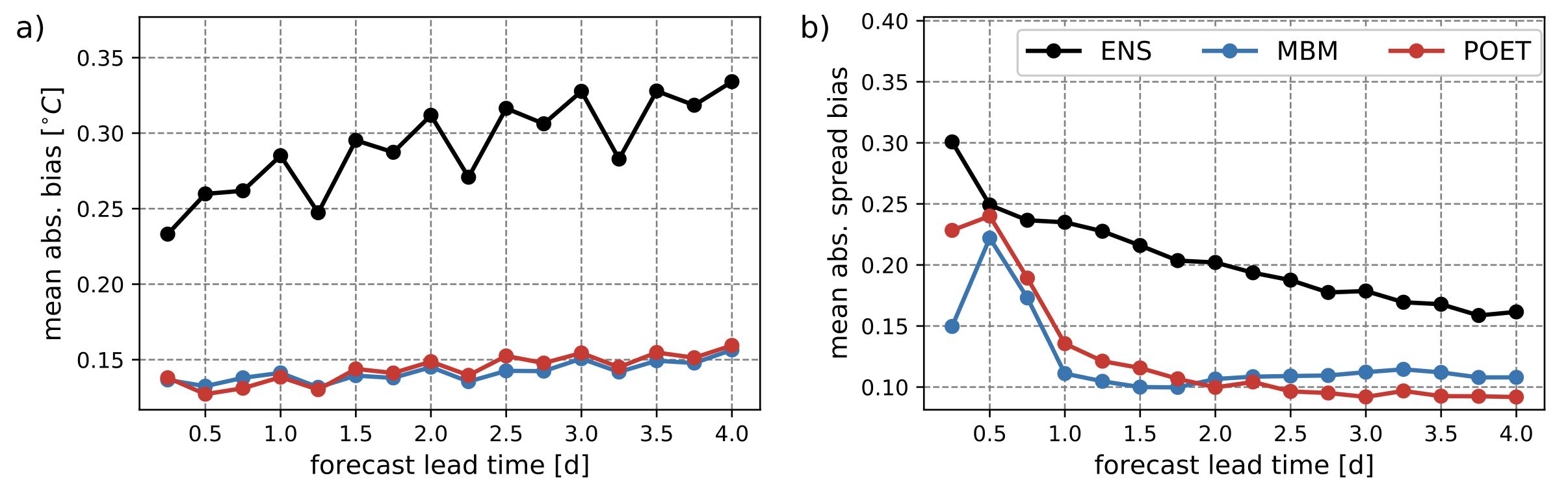}
    \caption{Mean absolute bias (a) and mean absolute spread bias (b) for 2m temperature forecasts. The closer to 0, the better.  }
    \label{fig:new_2t}
\end{figure}

\begin{figure}[ht]
    \centering
    \includegraphics[width=0.92\textwidth]{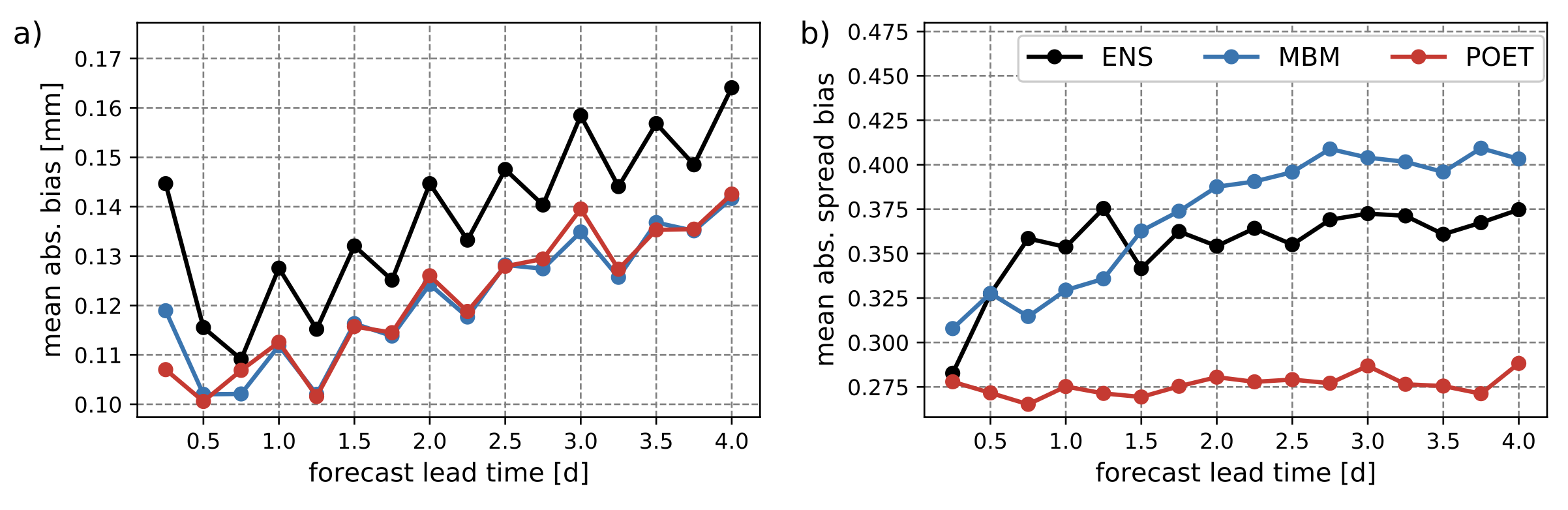}    
    \caption{Same as Fig. \ref{fig:new_2t} but for total precipitation.  }
    \label{fig:new_tp}
\end{figure}

Mean absolute bias and mean absolute spread bias for 2m temperature and precipitation are shown in Figs~\ref{fig:new_2t} and \ref{fig:new_tp}, respectively.

\subsubsection{Brier skill score}\label{sec:bssplots}
\begin{figure}[ht]
    \centering
    \includegraphics[width=0.92\textwidth]{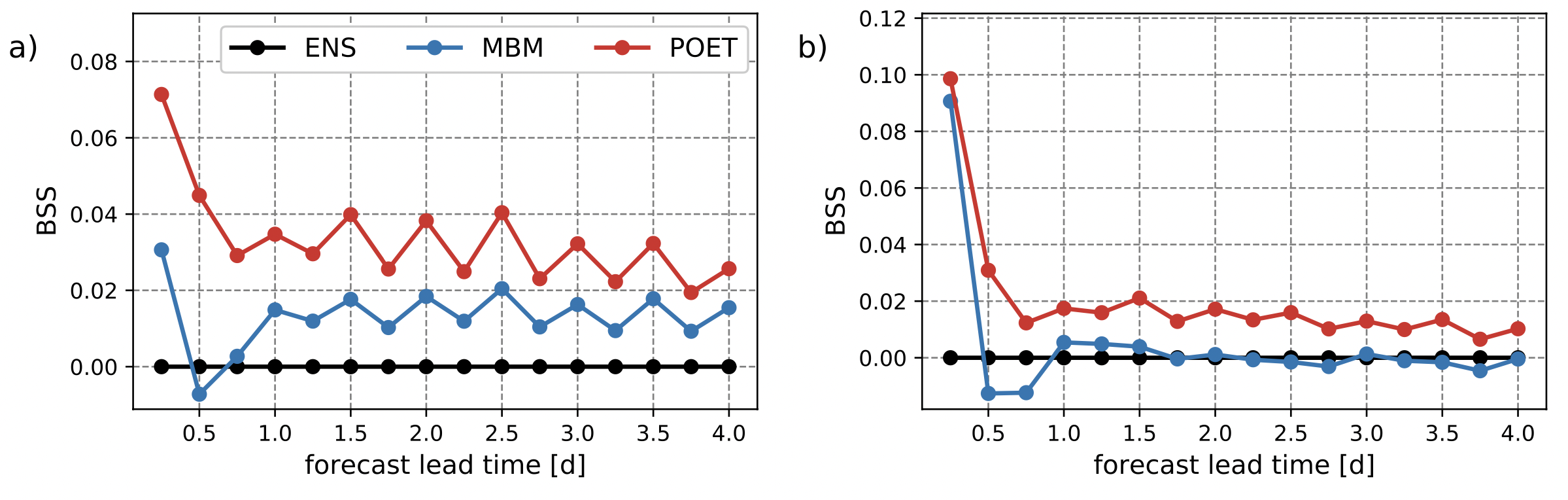}
    \caption{BSS (the larger, the better) for 2 threshold-exceeding events: (a) 1mm  and (b) 10mm for 3 ensemble forecasts: the raw ensemble, the MBM and the PoET-corrected forecasts. }
    \label{fig:brierscore}
\end{figure}

The Brier score (BS) is used to assessed the performance of a binary probability forecast for a given event \citep{brier50}. It is computed here to better understand the impact of post-processing on the results in terms of CRPS. Indeed, the CRPS corresponds to the integral of the Brier score over all possible events. Fig.~\ref{fig:brierscore} shows the Brier skill score for two precipitation events defined as precipitation exceeding 1mm and 10mm in 6h, respectively.

\end{document}